\documentclass[conference]{IEEEtran}
\usepackage{graphicx} 
\usepackage{url}
\usepackage{multirow}
\usepackage{hyperref}
\usepackage{tcolorbox}

\title{Towards green AI-based software systems: an architecture-centric approach (GAISSA)}
\author{\IEEEauthorblockN{Silverio Martínez-Fernández}
\IEEEauthorblockA{
\textit{Universitat Politècnica de Catalunya}\\
Barcelona, Catalunya, Spain\\
silverio.martinez@upc.edu}
\and
\IEEEauthorblockN{Xavier Franch}
\IEEEauthorblockA{
\textit{Universitat Politècnica de Catalunya}\\
Barcelona, Catalunya, Spain\\
xavier.franch@upc.edu}
\and
\IEEEauthorblockN{Francisco Durán}
\IEEEauthorblockA{
\textit{Universitat Politècnica de Catalunya}\\
Barcelona, Catalunya, Spain\\
francisco.duran.lopez@upc.edu}
}

\date{March 2023}

\begin{document}

\maketitle

\begin{abstract}
Nowadays, AI-based systems have achieved outstanding results and have outperformed humans in different domains. However, the processes of training AI models and inferring from them require high computational resources, which pose a significant challenge in the current energy efficiency societal demand. To cope with this challenge, this research project paper describes the main vision, goals, and expected outcomes of the GAISSA project. The GAISSA project aims at providing data scientists and software engineers tool-supported, architecture-centric methods for the modelling and development of green AI-based systems. Although the project is in an initial stage, we describe the current research results, which illustrate the potential to achieve GAISSA objectives.

\end{abstract}


\section{Introduction}

The use of AI-based systems have become widespread in modern society, requiring extremely high computational resources. Artificial Intelligence (AI) plays a key role in the world we live in. Nowadays, AI-based systems, defined as those software systems that integrate AI models and components \cite{martinez2022software}, have achieved outstanding results and have outperformed humans in areas as diverse as image processing, multi-object tracking, speech and facial recognition, automatic machine translation, character and handwriting recognition, etc., with applications in critical domains such as healthcare and autonomous driving \cite{deng2018artificial}. Key to the success of AI-based systems is the widespread adoption of Machine Learning (ML) and its subfamily Deep Learning (DL) approaches to AI. However, the processes of training AI models and inferring from them require high computational resources, especially considering the large amounts of data required. This requirement does not become a showstopper for AI-based systems’ adoption since we are in the middle of the Big Data, Big Compute, Big Models revolution, yet it creates a challenge in the current energy efficiency societal demand.

Large AI and software companies shall follow energy efficiency regulations. The EU has agreed ambitious targets for 2030 regarding greenhouse gas (GHG) emission reductions, renewable energy and energy efficiency, and has published its strategic long-term vision for a prosperous, competitive and climate-neutral economy by 2050 \cite{eu2019communication}. In this context, we can find European guidelines on reporting climate-related information for large non-financial companies \cite{eu2019communication}. These guidelines include Key Performance Indicators (KPI) for GHG emissions and energy, among others. Indeed, the EU 2020/852 Regulation \cite{european2020regulation} states: ``It is therefore appropriate to require the annual publication of such KPIs by such large companies and to further define that requirement in delegated acts''. Moving towards the ecological transition, such reporting is essential for software and AI companies to comply with energy efficiency regulations and therefore produce less data-intensive and lighter models. 

Current approaches to build AI-based systems mainly target accuracy, rarely addressing energy efficiency as well. All the investments in the modelling and development of highly accurate AI-based systems have led to a dramatic growth in required data volume, AI models’ size, and infrastructure capacity. The endless pursuit of achieving the highest possible accuracy has led to the exponential scaling of AI with significant energy and environmental footprint implications, which collides with the social benefits AI-based systems bring. According to Schwartz et al. \cite{schwartz2020green}, the computing resources used for training DL models has risen by a factor of 300.000x in only 6 years (2012-2018) due to modelling and algorithmic choices, over hardware considerations. \cite{schwartz2020green}. Consequences are dramatic to every single domain benefitting from AI-based systems. For instance, Wu et al. show that the carbon footprint of training one large ML model for autonomous vehicles is equivalent to 242,231 miles driven by an average passenger vehicle \cite{wu2022sustainable}. But this is only one aspect. To fully understand the real environmental impact, we must consider the AI ecosystem holistically going forward — beyond looking at model training alone and by accounting for both operational and embodied carbon footprint of AI \cite{wu2022sustainable}.

In the pursuit of green AI-based systems, the GAISSA project, which stands for towards Green AI-based Software Systems: an Architecture-centric approach, proposes the adoption of architecture-centric methods. For green AI we understand approaches to build AI models that are environmentally friendly and inclusive \cite{schwartz2020green}. In contrast, red AI refers to approaches to build AI models seeking to improve accuracy (or related measures) through the use of massive computational power while disregarding the cost \cite{schwartz2020green}. While the last few years have witnessed calls to act upon the huge energy consumption of AI-based systems, evidence shows that current architecture-centric methods do not aim to improve energy efficiency for such type of systems \cite{schwartz2020green, coeckelbergh2021ai, wu2022sustainable}. This is in contrast to current practices applied in traditional software systems’ development, where architectural decision-making has been reported as a principal approach when aiming at improving any quality-related aspect of a software system, including sustainability aspects \cite{stier2015model}.

\begin{tcolorbox}
\textbf{PROJECT HYPOTHESIS}: When modelling and developing green AI-based systems, the impact of architectural decisions on energy efficiency must be better understood, defined, reported, and managed in order to deliver AI-based systems with less demanding computational power needs.
\end{tcolorbox}

The paper proceeds as follows. Section \ref{state-of-the-art} reviews the state of the art. Section \ref{objectives} and Section \ref{methodology} present the objectives
of GAISSA and its methodology, respectively. Section \ref{research-team} presents the research team. Section \ref{expected-outputs} outlines the expected tangible outputs. Section \ref{current-results}
summarizes
the initial results of GAISSA. Finally, Section \ref{conclusions} draws the conclusions.

\section{State-of-the-art} \label{state-of-the-art}

In recent years, software sustainability has become a new research field \cite[ch.~1]{calero2015green}\cite[ch.~1]{calero2021software}, which in turn comprises three dimensions: environmental sustainability, human sustainability, and economic sustainability. GAISSA focuses on environmental sustainability, defined as ``how software product development, maintenance, and use affect energy consumption and the consumption of other natural resources. [...] This dimension is also known as Green Software'' \cite[ch.~1]{calero2015green}. For the sake of simplicity, in GAISSA we use the term {\bf green AI-based system} to refer to environmentally sustainable-aware AI-based systems. Also, we use the term {\bf greenability} defined as the ``degree to which a product lasts over time, optimising the parameters, the amounts of energy and the resources used'' \cite[ch.~10]{calero2015green}.

The subsequent state of the art discusses the current proposals to construct green AI-based systems, with focus on green-aware architecture-centric methods.

\subsection{Software measurement and quality models for green AI-based systems} 

The idea of considering sustainable software and software engineering alone without the hardware aspects was not fully recognized until 2010 \cite[ch.~6]{calero2015green}. Several quality models for the general software engineering domain that consider sustainability exist. Calero et al. \cite[ch.~10]{calero2015green} extend the ISO/IEC 25010 standard models for software product quality and quality in use with a new greenability quality characteristic that is further refined into several sub-characteristics: (i) energy efficiency, (ii) resource optimisation, (iii) capacity optimisation, (iv) perdurability. The above work is complemented by Moraga et al. \cite[ch.~11]{calero2015green} who have identified measures from the literature that can be used to evaluate the greenability characteristic of Calero et al.’s quality model and have pointed out that the number of measures for all the sub-characteristics is scarce. Additionally, Calero et al. propose a process to evaluate the software energy efficiency quality factor \cite[ch.~3]{calero2021software}.

Other approaches in this direction exist. For instance, Naumann et al. propose the GREENSOFT model for green and sustainable software and its engineering \cite{naumann2011greensoft}, and Taina et al. provide a layered model based on GREENSOFT to help to measure green software quality from different points of view \cite[ch.~6]{calero2015green}. All these previous proposals focus on the general software engineering domain and do not refer to the specific case of AI-based systems.

We think it is essential to distinguish the AI-based systems case from the extensive domain of software engineering, given its unique characteristics. AI-based systems have a number of specificities that need to be considered in a quality model for green AI. For instance, the ML and DL pipelines include specific phases such as data collection, model exploration and experimentation, model training and model optimization and, therefore, aspects such as the frequency of training and scale of each development phase, do matter \cite{wu2022sustainable}. Additionally, it is a challenge to measure the green qualities of AI software \cite{schwartz2020green}. Therefore, we argue that greenability should be added to the few existing proposals that target AI-based systems quality, such as the work by Siebert et al., who presented a systematic process to build quality models for these systems, but not targeting greenability \cite{siebert2022construction}.

\subsection{Architectural decisions and platforms for less energy-demanding AI modelling}

 Balancing the energy efficiency and accuracy of AI models is key to pursuing greener AI. However, evaluating and optimizing energy efficiency of AI models operating in new emerging scenarios, e.g. edge devices, is challenging \cite{castanyer2021integration}. Strubell et al. were pioneers in bringing the attention of AI researchers by quantifying the approximate environmental costs of training a variety of successful neural network models \cite{strubell2019energy}. While there is no community consensus on standards about which energy efficiency metrics to report, there have been some recent and scarce examples on computationally and energy-efficient AI models from the architectural perspective \cite{sanh2019distilbert}\cite{borgeaud2022improving}. Despite these first promising results, architecture-centric methods are still ad-hoc, emerging approaches in very specific projects that can hardly be considered transferable to other contexts. 
 
Furthermore, as highlighted by Cowls et al., there is a need to ``assess the carbon footprint of AI models that appear on popular libraries and platforms, such as PyTorch, TensorFlow and Hugging Face, to inform users about their environmental costs'' \cite{cowls2021ai}. This could help reason not only on the energy efficiency of training AI models, but also on the energy efficiency AI platforms where those are deployed. Deploying AI models to operate in environments with low computational resources constraints the design decisions of AI-based systems. One of such low computational resources’ ecosystems are mobile applications. With the growth of the mobile applications market, approaches on the deployment of neural networks in lightweight devices are becoming popular given their ability to provide great services to the user \cite{castanyer2021integration,castanyer2021design}. Despite the widespread adoption of these AI platforms, their energy efficiency exploration is really limited. A notable exception is a just accepted study on a limited comparison between PyTorch and TensorFlow \cite{georgiou2022green}. To sum up, more empirical studies are needed to understand and optimize the energy efficiency of AI platforms under different contexts (mainly, AI algorithm type to deploy).

\subsection{Software architectures and architectural decisions for green AI-based systems}

A number of approaches advocate the use of architectural or design patterns for dealing with sustainability issues. For instance, Chinenyeze and Liu explore in detail a kind of architectural pattern emerging from the use of the offloading technique in the context of mobile cloud applications \cite[ch.~8]{calero2021software}. At a more fine-grained level, Feitosa et al. present a set of 22 energy-aware design patterns emerging from the state of practice in mobile application design \cite{feitosa2021patterns}. However, only one of them (Enough Resolution) addresses energy consumption. None of these and similar sustainability-aware pattern-based approaches mention the particular case of AI-based systems, where sustainability issues are especially challenging.

Conversely, there are research works proposing the application of architectural and design patterns in the context of AI-based systems. Yokohama proposes a specific ``ML architectural pattern'' that keeps the business logic and the ML components separated \cite{yokoyama2019machine}. Architectural patterns for AI-based systems may be in an implicit form. For instance, Amershi et al. report a nine-stage workflow that captures the general approach to AI-based systems design at Microsoft Research \cite{amershi2019software}. At the level of design patterns, Washizaki et al. identify and report practical usage of a set of 15 design patterns for AI-based systems \cite{washizaki2022software}. Design patterns may be used also not for developing from scratch but to refactor an existing system’s architecture, as proposed by Ribeiro et al. \cite{ribeiro2019microservice}. Again, none of these pattern-based approaches for AI-based systems address greenability issues, but only other criteria such as prediction performance, scalability, maintenance and reusability \cite{ribeiro2019microservice}. 

Other architecture-centric works explore sustainability issues without using patterns as instrument. We may mention García-Rodriguez et al., who propose to integrate classical with green maintenance to become software systems developed following a classical approach into green software systems. Moreover, they analyse to what extent refactoring classic software flaws has an impact on software greenability and define the concept of ecological debt \cite[ch.~9]{calero2021software}. Although the proposal focused on software systems in general, the ideas and results may be refined to the context of AI-based systems and the measure of technical debt could be eventually integrated as information in an envisioned set of sustainability-aware patterns for AI-based systems, complementing existing works that measure technical debt for AI-based systems without considering greenability \cite{bogner2021characterizing, lenarduzzi2021systematic}.

\subsection{Tools and techniques for the modelling and development of green AI-based systems}

Anwar et al. analyse and report 24 support tools that help developers build energy-efficient Android applications during the development and maintenance phases \cite[ch.~7]{calero2021software}. These tools are classified into three categories: Profilers, Detectors, and Optimizers. A Profiler is a software tool that measures the energy consumption of a given application. A Detector is a tool that ``only'' identifies and detects so-called ``energy bugs and code smells.'' Finally, an Optimizer is a tool that not only identifies ``energy bugs and code smells'' but also fixes them by refactoring source code to improve energy consumption. 

Raturi et al. \cite[ch.~2]{calero2015green} shift the attention to profiling the energy consumption in software engineering environments. In particular, they present the Joulery Energy Dashboard that gathers, aggregates, and visualizes energy data from networked devices in a software engineering environment to identify energy sinks that can then be traced back to the different conducted engineering activities.

Lannelongue et al. \cite{lannelongue2021green} present a tool for estimating computation's carbon footprint by considering several sources of energy consumption, such as the runtime, the number and type of the processor cores, the amount of memory requested, and the overhead of computing facilities and their geographic location. 

Recently, major cloud providers have released or announced the incoming release of dashboard tools to allow their customers to measure, visualize, report, and even reduce the carbon footprint of their cloud projects: Microsoft’s Emissions Impact Dashboard, Google’s Carbon Footprint, and Amazon’s AWS Customer Carbon Footprint Tool. 

However, no current tool supports the modelling and development cycles of engineering AI-based systems from an energy efficiency-aware perspective, providing assessments, estimations and what-if analysis of the energy consumption and carbon footprint for the different alternatives available in using third-party libraries, training AI models, or selecting deployment AI architectures and platforms.

\subsection{Research gap}\label{research-gap}

Although in the last few years the need of delivering greener AI-based systems has been increasingly recognized, current architecture-centric methods for green AI-based systems modelling and development are in their infancy. In order to understand, define, report and manage the energy efficiency of the architectural decisions over AI-based systems, we summarize the identified gaps emerging from the state of the art conducted in this section:
\begin{itemize}

\item [\textbf{G1:}] A quality model for software greenability evaluation of AI-based systems.
\item [\textbf{G2:}] Predictive models and context-aware evaluation of AI platforms for optimizing the energy efficiency of AI models’ training and deployment.
\item [\textbf{G3:}] Architecture and design patterns for developing green AI-based systems and deploying them into their operational environment.
\item [\textbf{G4:}] Analytic tools and assets to support analysis, decision-making and reporting during the modelling and development of green AI-based systems.
\end{itemize}

\section{Objectives}\label{objectives}

The project's main objective is:

\begin{tcolorbox}
\textbf{To provide data scientists and software engineers tool-supported, architecture-centric methods for the modelling and development of green AI-based systems}.
\end{tcolorbox}

It further decomposes into four specific objectives (\textbf{O1-O4}) addressing the four research gaps (\textbf{G1-G4}) identified in \ref{research-gap}:

\begin{itemize}
\item [\textbf{O1:}] \textbf{Define, implement, and evaluate a quality model for measuring the greenability of AI-based systems.} This objective sets up the basis of AI-based systems’ greenability with: (i) a hierarchical structure of greenability-related quality factors, from abstract green quality indicators down to concrete greenability metrics; (ii) operationalized metrics for all these quality factors; (iii) understanding of the synergies and conflicts among greenability quality factors and other qualities, such as accuracy of results or data privacy. 
\item [\textbf{O2:}] \textbf{Define, implement and evaluate architecture-centric methods to guide the training and deployment of green AI models and measure energy efficiency of AI platforms.} This objective provides data scientists with: (i) predictive models to understand how design decisions on AI models affect energy efficiency, in order to train sustainability-aware greener AI models; (ii) context-aware measurement of the energy consumption of AI platforms, to make informed decisions about where to deploy AI models.
\item [\textbf{O3:}] \textbf{Define, implement and evaluate architecture-centric methods to guide the development of green AI-based systems.} This objective provides software engineers with: (i) the identification of architectural concerns related to the quality factors mentioned in O1, which act as drivers for the architectural design of AI-based systems; (ii) a catalogue of architecture and design patterns to build green AI-based systems with the description of their forces (quality factors to optimize) and how they interact/conflict with each other. 
\item [\textbf{O4:}] \textbf{Define, implement and evaluate analytic tools and assets to support greenability-driven analysis, decision-making and reporting during the modelling and development of AI models and AI-based systems.} This objective provides data scientists and software engineers with tools and assets to (i) yield visualizations, explanations and reports of greenability measurements and estimations; (ii) enable predictions, what-if analysis and comparison of alternatives to target greenability-related outcomes.
\end{itemize}

\begin{figure*}[!ht]
    \center
    \includegraphics[width=\textwidth]{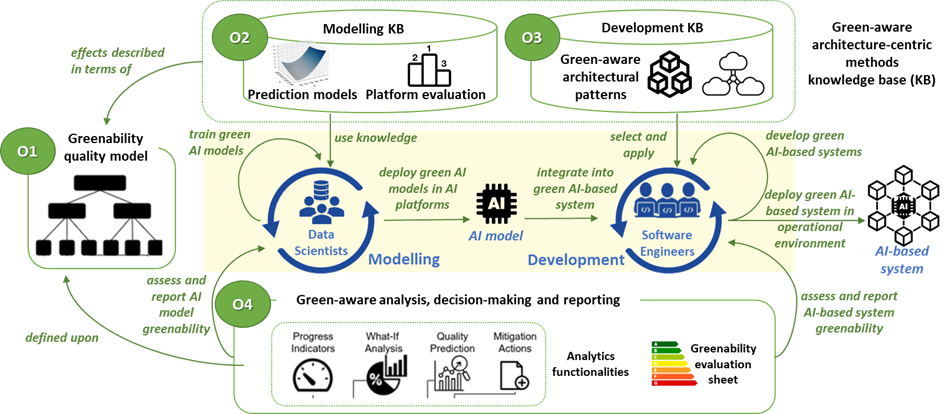}
    \caption{\label{schema}The GAISSA approach for green AI modelling and green AI-based system development.}
\end{figure*}

Figure \ref{schema} depicts the GAISSA approach, including the main activities and assets related to GAISSA objectives, O1–O4 (detailed above). The project is framed in the context of the AI engineering lifecycle defined by Lwatakare et al. \cite{lwakatare2020devops}, which consists of four stages also organized as cycles: Data Management, Modelling, Development and Operation. GAISSA focuses on the two middle stages: Modelling, in which data scientists train the AI model for its later deployment into AI platforms; and Development, in which software engineers integrate the AI model into the software, and develop and deploy the AI-based system. These two stages face challenges that go beyond those in traditional software engineering \cite{martinez2022software}. For instance, all the experimentation and training done during Modelling is much more time consuming than in traditional software systems. Scoping the project in these two stages allows us to define ambitious, impactful and realistic objectives for a two-year period.

Furthermore, we may say that GAISSA is an interdisciplinary project. It includes aspects of various disciplines: (i) Artificial Intelligence, with focus on green AI; (ii) Software Engineering, with focus on software architectures and software quality (through quality models and analytics); (iii) Sustainability, with focus on energy efficiency and ecological transition on ICT.

\section{Methodology}\label{methodology}

The project is organized into four phases, with the second and third phases conducted according to agile principles (\textbf{Mk} stands as abbreviation of Month k from the start of the project execution):
\begin{enumerate}

\item \textbf{Project start (M1-M2):} initial state-of-the-art performed; development environment set up; initial evaluation protocol completed; dissemination, communication, exploitation \& management started.
\item \textbf{Proof-of-concept (M3-M12):} deployment of initial set of implemented and evaluated techniques; emphasis on the individual results from scientific WPs (e.g., quality model, architecture-centric methods) rather than on the final integrated complete system.
\item \textbf{Integrated system (M13-M22):} portfolio of quality model and architecture-centric methods, evaluated and integrated techniques showing the full functionality and potential; emphasis shifted to the integrated tool to manage energy efficiency of AI-based systems for software engineers and data scientists, and the exploitation of the greenability evaluation sheet.
\item \textbf{Project finalization and sustainability (M23-M24):} finalization of evaluation; impact-related final actions, packaging of results and code under an open science approach.
\end{enumerate}

The overall strategy of the work plan is organised into the following Work Packages (\textbf{WP}) and areas:

\begin{enumerate}
\item \textbf{Scientific Core (WP1-WP3):} formulates the scientific methods and techniques produced in the project; each of these WP1-WP3 is bound to the corresponding project objective O1-O3.
\begin{itemize}
\item  [\textbf{WP1:}] A quality model for greenability of AI-based systems (O1).
\item  [\textbf{WP2:}] Architectural-centric methods for greener AI modelling (O2).
\item  [\textbf{WP3:}] Architectural-centric methods for greener AI development (O3).
\end{itemize}

\item \textbf{Operationalization (WP4):} implements the individual tools that implement the methods and techniques defined in WP1-WP3, and designs and implements analytic tools and assets to assist green-aware decision-making. Therefore, it covers not only O4 but also the implementation part of O1-O3.

\begin{itemize}
\item  [\textbf{WP4:}] Tools and techniques for analysis, decision-making and reporting for green AI-based systems (O1, O2, O3, O4).
\end{itemize}

\item \textbf{Evaluation (WP5):} sets up an empirically-based evaluation plan to consolidate the results from the scientific core (WP1-WP3) and developed tools (WP4), covering thus the evaluation part of O1-O4. 

\begin{itemize}
\item  [\textbf{WP5:}] Evaluation (O1, O2, O3, O4).
\end{itemize}

\item \textbf{Project-Wide Activities (WP6-WP7):} which include traversal activities that involve all the other WPs.

\begin{itemize}
\item  [\textbf{WP6:}] Dissemination, Communication and Exploitation.
\item  [\textbf{WP7:}] Project Management.
\end{itemize}

\end{enumerate}

\section{Research team}\label{research-team}
The GAISSA project is being executed by the Software and Service Engineering research group at the UPC (\url{https://gessi.upc.edu/en}). Being a strategic project for the group, all nine GESSI senior researchers are participating in its Research Team (RT). In addition, a number of PhD, MSc and BSc students are undertaking research and development activities in the context of GAISSA, some of them being hired specifically for the project.
In order to improve the quality of our research, we have involved other researchers in the project in two different ways (see Table \ref{table-team-1}):
\begin{itemize}
\item We collaborate in specific tasks with five experienced researchers from top-tier European universities working in the project areas.

\item We include an interdisciplinary Advisory Board (AB) composed of six world-wide experts in the project areas who will critically assess the progress of the project and propose research directions. The AB members cover the different areas of the project, supporting interdisciplinarity and therefore helping to identify synergies, analogies, paradoxes and approaches from multiple viewpoints. 
\end{itemize}

\begin{table}[htbp]
\caption{GAISSA external team}
\label{table-team-1}
\begin{center}
\scalebox{0.9}{
\begin{tabular}{ll}
\hline
\multicolumn{1}{|c|}{\textbf{Role}}                                                    & \multicolumn{1}{c|}{\textbf{Members}}                                                                                                                                                                                                                                                                                           \\ \hline
\multicolumn{1}{|l|}{\begin{tabular}[c]{@{}l@{}}Abroad\\ collaborators\end{tabular}}   & \multicolumn{1}{l|}{\begin{tabular}[c]{@{}l@{}}Justus Bogner (U. Stuttgart), \\ Luís   Cruz (U. Delft), \\ Filippo Lanubile (U. Bari),\\ Valentina   Lenarduzzi (U. Oulu),\\ Roberto Verdecchia (Vrije U. Amsterdam)\end{tabular}}                                                                                              \\ \hline
\multicolumn{1}{|l|}{\begin{tabular}[c]{@{}l@{}}Advisory\\ Board {[}*{]}\end{tabular}} & \multicolumn{1}{l|}{\begin{tabular}[c]{@{}l@{}}Schahram Dustdar (Technical U. Vienna; DE),\\ Andreas Jedlitschka (IESE Fraunhofer; DS, AI), \\ Patricia Lago (Vrije U. Amsterdam; SA, GR), \\ Grace Lewis (SEI-CMU; SA, AI), \\ Birgit Penzenstadler (Chalmers U.; GR, SU), \\ Monica della Pirriera (Leitat; SU)\end{tabular}} \\ \hline
\multicolumn{2}{l}{\begin{tabular}[c]{@{}l@{}}$*$Initials for expertise areas: SU, sustainability; GR: greenability; AI, artificial\\ intelligence; SA, software architecture; DE, advanced deployment models (edge\\ computing, fog computing, …); DS: data science.\end{tabular}}                                                                                                                                      
\end{tabular}}
\end{center}
\end{table}

\section{Expected outputs}\label{expected-outputs}

The main assets to be produced in the project that can be individually transferred are:

\begin{enumerate}    
  \item \textbf{Greenability characterization and measurement.} Produced in WP1. Comprises:
  \begin{enumerate}
    \item Greenability quality model (indicators, factors and metrics) for AI models and AI-based systems.
    \item Greenability evaluation sheet, including energy efficiency levels definition.
  \end{enumerate}
  
  \item \textbf{Architecture-centric support for green AI modelling.} Produced in WP2. Comprises:

  \begin{enumerate}
    \item Predictive models to reason on architecture-centric greenability of AI models.
    \item A context-aware evaluation of AI platforms’ energy efficiency for optimal AI models’ deployment.
  \end{enumerate}
  
  \item \textbf{Architecture-centric support for green AI-based system development.} Produced in WP3. Comprises:

  \begin{enumerate}
    \item A catalogue of architectural patterns for green AI-based systems.
    \item A catalogue of design patterns for green AI-based systems.
  \end{enumerate}
  
  \item \textbf{Tooling.} Produced in WP4. Comprises:

  \begin{enumerate}
    \item Loosely-coupled tools (in the form of microservices) supporting the project objectives. \label{outcomes-tooling-a}
    \item A visualisation and reporting tool for the domain of green AI-based systems, using (\ref{outcomes-tooling-a}) as needed.
    \item A software analytic tool for the domain of green AI-based systems, using (\ref{outcomes-tooling-a}) as needed.
    
  \end{enumerate}
  
  \item \textbf{Evaluation.} Produced in WP5. Comprises:

  \begin{enumerate}
    \item An open repository of studied AI-based use cases and their greenability evaluation sheets.
  \end{enumerate}

\end{enumerate}

\section{Current results}\label{current-results}

GAISSA started by December 2022. Despite being in its initial stage, the project has delivered a number of results that can be summarized by the following research activities:

Regarding O1, Martinez et al. \cite{martinez2022energy} considered the energy impact of repairing bugs automatically using APR (Automated Program Repair), which defines the foundations for measuring the energy consumption of APR activity. This study shows the existing trade-off between energy consumption and the ability to correctly repair bugs from point of view of Software Developers. Furthermore, this work confirms the importance of defining concrete greenability metrics and methods, and provides evidence that architecture-centric decisions in the development of software systems (program repair tools) differ on energy consumption.

As part of our research towards O2, Castanyer et al. \cite{castanyer2021integration} provided an analysis of the challenges found when developing AI mobile applications and analyzed different configurations of CNN models for the problem of optimizing the performance trade-off between accuracy and complexity of DL models in the same context of AI mobile applications. In a separate study, Castanyer et al. \cite{castanyer2021design} illustrated the relation that exists between design decisions, such as dataset, number of parameters, architecture type, and the performance of AI mobile applications. Furthermore, they provided insights on the AI mobile application development and on the use of profiling tools to monitor those applications.

Xu et al. \cite{xu2023energy} carried out an empirical study about the environmental impact in terms of energy consumed and $CO_2$ emissions, produced by using different CNN model architectures and infrastructure locations, the trade-off between accuracy and energy efficiency,  and a comparison of tools to measure the energy efficiency.
For their part, Durán et al. \cite{duran2022guiding} conducted an empirical study about retraining methods for CNN models against adversarial inputs. This study shows that architecture-centric decisions in a retraining stage of the models, such as retraining strategy and dataset configuration, have an impact not only on the accuracy of the models but on metrics of greenability-related quality factors such as execution time and resource utilization (number of inputs used for retraining).
Both studies are a step towards defining the relation between design decisions on AI models and their energy efficiency, and understanding the synergies between greenability-related quality factors and the most targeted metric in AI models literature, accuracy. In addition to that, the comparison of tools from the former study \cite{xu2023energy} is an important building block in our journey towards parameterizing the greenability measures from the quality model.

In their research, Del Rey et al. \cite{delrey2023training} measured the energy consumption of five DL model architectures and three training environments to analyze how these two architecture-centric decisions during development can affect the energy of training DL models. They showed that when doing a proper selection, the energy consumption can be reduced without diminishing the correctness of the model.

In relation to O3, Lanubile et al. \cite{lanubile2023teaching} proposed a project-based learning approach to teach MLOps, in order to provide both undergraduate and graduate students with theoretical and practical knowledge on building high-quality, production-grade ML
components, ready to be integrated into ML-enabled systems. This project-based course covers the full end-to-end ML component life cycle and has a clear milestone on sensitizing Higher Education students to build energy-efficient ML models.

From a methodological perspective, Franch et al. \cite{franch2022quatic} propose a preliminary ontology for architectural decision making. The ontology revolves around the traditional concept of Architectural Decision, which are of a particular Decision Type and are made in a specific Context. Architectural decisions affect a number of Architectural Elements, which can be AI-related Architectural Elements, and they Impact (either positively or negatively) to one or more Quality Attributes. In another work, Franch et al. \cite{franch2023refsq} reflect on the role of the requirements engineer in the development of AI-based systems and among other things, it establishes the responsibilities of this and other roles, which affects the software architect who may be highly involved in the process of defining and assessing software requirements.

Ahmed et al. \cite{ahmedevolution} conducted an empirical study to evaluate the energy consumption during the evolution of three Kotlin mobile apps and identified a growing trend with different possible causes, which are related to OS upgrades, new features, poorly chosen design patterns and libraries, UI issues, and unstable app versions. This study indicates that architecture-centric decisions during the evolution over time of a software have a considerable influence on the consumed energy.

By their part, Del Rey et al. \cite{delrey2023review} provided a brief overview on 20 recent studies focused on energy efficiency when deploying ML models on the edge. They made a classification of four main themes according to their main contributions. Furthermore, they identified that most of the work is focused on improving energy efficiency by optimizing the edge devices' workload and communication, and that there is scarce research on understanding the factors that impact the energy consumption and carbon footprint. This review provides deeper insight into the state of the art of ML models and systems in green deployment for Edge AI, helping researchers to understand the current problems on that topic.

With respect to O4, Castaño et al. \cite{castano2023exploring} did an exploratory mining study of ML models from the Hugging Face Hub, they analyzed how carbon emissions are measured and reported, and studied the model attributes that impact the carbon emissions. They found a lack of awareness of green AI by the Hugging Face community and uncovered correlations between the carbon emissions and model attributes such as model size, dataset size and ML application domain. In addition to that, in order to increase the green-aware ML model development, they also propose two categorization for the models: based on their carbon emission reporting practices and based on their carbon efficiency. Furthermore, they propose guidelines to standardize reporting of carbon emissions. This serves as a crucial step in the pursuit of classifying ML models according to their energy efficiency and increasing the ML sustainability awareness.

\section{Conclusions}\label{conclusions}

In this paper we have presented the two-year (December 2022-November 2024) GAISSA project. This project will provide software engineers and data scientists with the guidelines for the modelling and development of green AI-based systems. Furthermore, the proposed GAISSA project has the potential to make a significant environmental impact and to provide a paradigm switch in which data scientists and software engineers build their AI-based systems more energy efficient. This results in saving energy, bringing social advantages by mitigating global climate change, and economic advantages by obtaining cost savings. Up to this point, the project tasks are being developed as expected and the first results bring us closer to meeting GAISSA goals. More information is available on the project website, \url{https://gaissa.upc.edu/en}.

\section*{Acknowledgment}

This paper has been funded by the GAISSA Spanish research project (ref. TED2021-130923B-I00; MCIN/AEI/10.13039/501100011033).

\bibliographystyle{IEEEtran}
\bibliography{bibliography}

\end{document}